\documentclass{aa}  

\usepackage{graphicx}
\usepackage{subfigure} 
\usepackage{footnote}
\makesavenoteenv{tabular}
\makesavenoteenv{table}
\usepackage{txfonts}
\usepackage{natbib}
\usepackage{hyperref}
\usepackage{float}
\usepackage{mathtools}
\usepackage{dcolumn}
\usepackage{placeins}

\begin{document}

   \title{A deep narrowband survey for planetary nebulae at the outskirts of M33 \thanks{Based on observations made with the Isaac Newton Telescope and service observations made with the William Herschel Telescope operated on the island of La Palma by the Isaac Newton Group of Telescopes in the Spanish Observatorio del Roque de los Muchachos of the Instituto de Astrofísica de Canarias.}}


   \author{R. Galera-Rosillo
          \inst{1,2},
          R.L.M. Corradi
          \inst{3,1,2},
          A. Mampaso
          \inst{1,2}
          }        

\institute{Instituto de Astrof\'isica de Canarias, Calle V\'ia L\'actea, s/n, E-38205, La Laguna, Tenerife, Spain\\             
\and
Departamento de Astrof\'isica, Universidad de La Laguna, E-38206, La Laguna, Tenerife, Spain\\           
\and
GRANTECAN, Cuesta de San José s/n, E-38712 , Breña Baja, La Palma, Spain\\
\\
\email{rgr@iac.es}}


  \abstract  
{Planetary nebulae (PNe) are excellent tracers of stellar populations with low surface brightness, and therefore provide a powerful method to detect and explore the rich system of substructures discovered around the main spiral galaxies of the Local Group.}
{We searched the outskirts of the Local Group spiral galaxy M33 (the Triangulum) for PNe to gain new insights into the extended stellar substructure on the northern side of the disc and to study the existence of a faint classical halo.} 
{The search is based on wide field imaging covering a 4.5 square degree area out to a maximum projected distance of about 40 kpc from the centre of the galaxy. The PN candidates are detected by the combination of images obtained in narrowband filters selecting the [OIII]$\lambda5007\AA$ and H$\alpha$ + [NII] nebular lines and in the continuum g' and r' broadband filters.}
{Inside the bright optical disc of M33, eight new PN candidates were identified, three of which were spectroscopically confirmed. No PN candidates were found outside the limits of the disc. Fourteen additional sources showing [OIII] excess were also discovered. } 
{The absence of bright PN candidates in the area outside the galaxy disc covered by this survey sets an upper limit to the luminosity of the underlying population of $\mathrm{\sim1.6\cdot10^{7}L_{\odot}}$, suggesting the lack of a massive classical halo, which is in agreement with the results obtained using the RGB population.}

   \keywords{ISM: planetary nebulae: general - galaxies: individual (M33, M31) - galaxies: interactions - galaxies: outskirts
   }

   \maketitle


\section{Introduction}

Planetary nebulae (PNe) are widely used to trace stellar populations in external galaxies, providing accurate information regarding the luminosity, age, metallicity, and dynamics of their progenitor stellar population. Their bright emission lines, and in particular the  [OIII]$\mathrm{\lambda5007\AA}$ nebular line, provide a distinctive characteristic that makes them easy to identify in continuum-subtracted, narrowband images. Planetary nebulae are especially useful to highlight stellar population in regions of very low surface brightness, such as the outermost regions of galaxies \citep{Corradi2015} and the intergalactic medium (\citealt{Longobardi2015}, \citealt{Arnaboldi2016IAUS}) as far as 100 Mpc (e.g. \citealt{Gerhard2007}).

The rich and complex structures highlighted in recent years in numerous galaxies by extensive broadband imaging surveys  have opened a unique laboratory to test not only galactic evolution models but also the history and possible interactions between galaxies in and outside the Local Group. In this respect, PNe can play a key role as chemical and dynamical witnesses of these processes.\\ Assuming a single-star evolutionary scenario, PN progenitors are expected to be low and intermediate mass stars 
with masses between $\sim$0.7 and 8M$\odot$ \citep{Badenes2015}. During their evolution these stars do not significantly modify abundances of $\alpha$-elements such as oxygen, neon, argon, and sulphur,  thereby preserving the original composition of the gas cloud from which the progenitor star formed. Because of their typical progenitor star lifetimes of between 0.1 to 10 Gyr,  PNe offer a direct look into the past \citep{Magrini2009}. These objects are used to determine metallicity gradients and their temporal evolution by means of comparing the PNe metallicity with those of younger stellar populations, such as HII regions or OB stars (\citealt{MagriniGoncalves2009}, \citealt{Annibali2017}). 

The kinematics of the  stellar population can be inferred by measuring the line-of-sight velocity of a PN population. (\citealt{Arnaboldi2012IAUS} and references therein).
Moreover, the [OIII]$\lambda$5007$\AA$ PN luminosity function (PNLF) has been proven to be a robust secondary extragalactic distance candle, providing 10$\%$ accuracy in large galaxies of all Hubble types \citep{Ciardullo2013IAUS}.

Deep studies of the bright disc of M33 confirmed a wealth of HII regions (\citealt{Wyder1997}, \citealt{Hodge1999Cat}) and PNe. The first comprehensive search for PNe was performed by \cite{Magrini2000} and provided 131 PN candidates. A subsequent search by \cite{Ciardullo2004} added several new candidates and spectroscopically confirmed 140 PNe. The analysis of the spectra pointed out dynamical properties that are consistent with the old disc from the velocity dispersion field of 138 of the PNe; the other 2 PNe are possibly related to a  spheroidal component. A detailed discussion of the PNLF of M33 was included in this analysis along with a comparison of the inner and outer disc PNLF, thereby supporting the independence of the bright cut-off with the metallicity suggested in \citet{Ciardullo2002} and displaying a decrease of the PNLF at fainter magnitudes.

The radial metallicity gradient of M33 and its evolution with time was inferred by \citet{Magrini2009} and revisited in \citet{Magrini2016}; these works claimed a tight relationship between O/H and Ne/H abundances and excluded the modification of both elements by the PN progenitors. The comparison between PNe and HII regions abundances homogeneously analysed in \citet{Magrini2016} indicates a low global enrichment of the M33 disc from the epoch of the formation of the PN progenitors to the present time, following the nearly identical slope of the oxygen gradients of PNe and HII regions.

The PN radial gradient data extend out to 8 Kpc from the centre of the galaxy. It would be very  interesting to explore whether these measurements can be extended to larger galactocentric distances and to compare these measurements with results obtained with broadband stellar studies of the outskirts of M33 and test whether a flattening of the gradient is present as found in similar galaxies (e.g. \citealt{Bresolin2009}).
Motivated by the paradigmatic case of M31, which shows a variety of organized and irregular structures located well beyond its classical disc, and inside which a number of PNe have been found \citep{Merrett2006}, we decided to embark on the search for analogous structures in M33 using PNe.

Early observations of M33  in the 21 cm HI line \citep{Rogstad1976} showed a severe disc warp; this indicates that M33 is not as pristine as it appeared in the optical counterpart known at that time. A first deep optical survey \citep{Ferguson2007} showed no distortions or substructures on the outskirts of the galaxy, but subsequent and deeper surveys claimed, first, the detection of a stellar halo \citep{Ibata2007}, and then the presence of a  large, irregular, low surface brightness stellar substructure surrounding the  galaxy (\citeauthor{McConnachie2009} \citeyear{McConnachie2009}, \citeyear{2009NatureMcCo}, \citeyear{McConnachie2010}) .

\citet{McConnachie2009} found the structure to be composed of an old stellar population ($\mathrm{\langle[Fe/H]\rangle\sim-1.6}$ dex), reaching as far as about 40 kpc from the centre in the northwest and southeast quadrants. These authors also suggested that the tidal disruption of M33 in its orbit around M31 would be the cause of the formation of the structure. Further insights about this possible interaction between the two spiral galaxies were gained by means of N-body simulations \citep{2009NatureMcCo}. 
The interaction between the two galaxies would have occurred 2-3 Gyr ago and would have triggered a burst of star formation in both galaxies supported by studies of stellar populations located well beyond the limit of their bright classical discs (\citealt{Barker2011}, \citealt{Bernard2012}). 
More recent simulations, however, do not support this scenario \citep{Patel2017}.  

The PNe in M33 can be used as an independent tracer of the interaction and star forming episodes, as their brightest representatives are predicted to be produced by stars in an age range covering the period of the supposed M31-M33 encounter \citep{Schonberner2007}.
The presence, or otherwise, of PNe in these regions therefore provides key information about and a quantitative measure of the luminosity of the underlying stellar population \citep{Buzzoni2006}.


\section{Observations}

Images were taken during seven nights on September and October 2013 using the Wide Field Camera (WFC) at the prime focus of the 2.54 m Isaac Newton Telescope (INT) of the Observatorio del Roque de los Muchachos at the island of La Palma (Spain). 

The WFC is an optical mosaic camera consisting of four thinned EEV CCDs with 2048 x 4096 pixels each and a scale of 0.33 arcsec/pixel.
The size of the mosaic as projected in the sky is 34.2 arcmins with $\sim$1 arcmin inter-chip gaps.  

\begin{figure}[!h]
\centering
\includegraphics[width=\hsize]{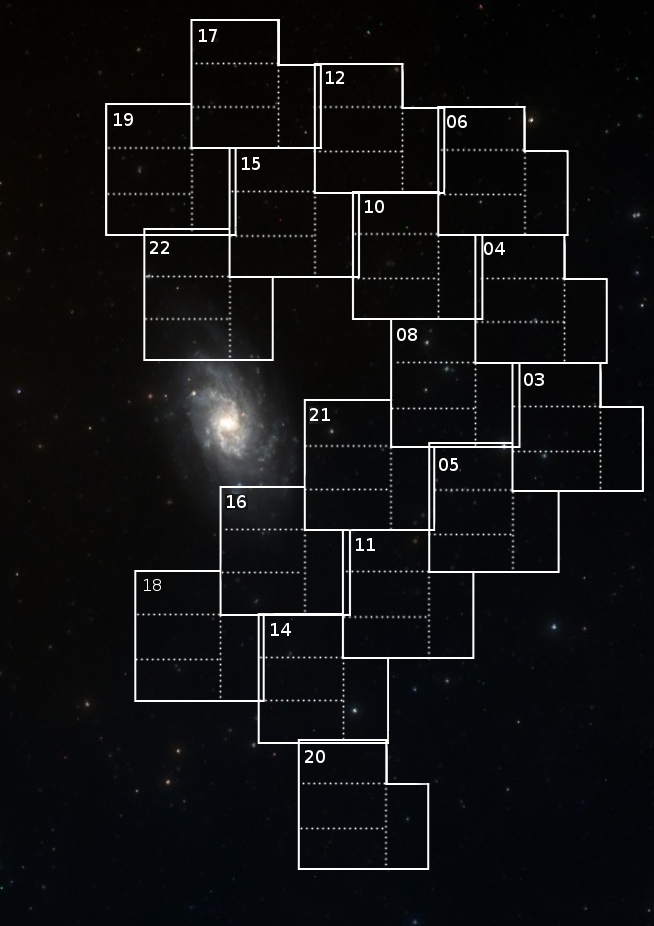}
\caption{M33 observed fields. The total field of view of the image is 3$^{\circ}$ x 4 $^{\circ}$. North is up, east at left.
}
\label{apuntados}
\end{figure}

Seventeen overlapping fields were observed (Fig.~\ref{apuntados}), covering a total region of 4.5 square degrees and reaching a maximum projected distance of about 40 kpc from the centre of the galaxy.
  
The emission-line filters used were [OIII], which has a central wavelength and full width at half maximum\ (FWHM) of 5008 and 100 $\AA,$ respectively; H$\alpha$+[NII] (6568/95 $\AA$); and the broadband g' (4846/1285 $\AA$) and r' (6240/1347 $\AA$) filters of the Sloan photometric system to evaluate the continuum emission and to distinguish PNe from stars. 
To allow cosmic ray removal, three exposures were obtained at each telescope position and for each filter, except at one of the telescope positions (M33 17) where only [OIII] and g' images could be taken. Exposure times were 1200 seconds for both [OIII] and H$\alpha$, and 400 seconds for r' and g'.
The seeing during observations ranged from 0.9 to 1.5 arcsec. The observing log is shown in Table~\ref{table:1}. 

Spectra of three PNe candidates were obtained in December 2014 in service-queue mode using the auxiliary-port camera and spectrograph ACAM at the 4.2m William Herschel Telescope (WHT), also at La Palma. 
The long slit was opened to  1$"$ and the exposure time for each object was $\sim$1 hour with the low resolution V400 grism, covering the wavelength range between 3500-9400 \AA. The log of these observations is presented in Table \ref{table:2}. 
Objects presented here are named GCMnn. 


\begin{table*}
\caption{INT imaging.}             
\label{table:1}      
\centering                          
\begin{tabular}{c c c c c c}        
\hline\hline                 
Field & R.A. J2000 & DEC J2000 & Date  \\    
\hline                        
M33 03 & 01:27:16.75    & +30:38:22.8 &  2013-10-11\\
M33 04 & 01:27:59.87    & +31:11:51.5 &  2013-10-12 \\ 
M33 05 & 01:28:58.07    & +30:17:46.2 &  2013-10-11 \\ 
M33 06 & 01:28:44.83    & +31:44:57.9 &  2013-10-12 \\ 
M33 08 & 01:29:43.66    & +30:50:06.8 &  2013-09-29 \\ 
M33 10 & 01:30:29.06    & +31:23:11.5 &  2013-09-29 \\
M33 11 & 01:30:41.39    & +29:55:13.3 &  2013-10-10 \\ 
M33 12 & 01:31:15.01    & +31:56:14.6 &  2013-10-13 \\
M33 14 & 01:32:23.25    & +29:33:20.4 &  2013-10-10 \\ 
M33 15 & 01:32:58.87    & +31:34:20.9 &  2013-09-28\\
M33 16 & 01:33:09.00    & +30:06:21.1 &  2013-10-09 \\
M33 18 & 01:34:50.47    & +29:44:21.5 &  2013-10-09 \\ 
M33 19 & 01:35:29.30    & +31:45:18.9 &  2013-09-28\\
M33 20 & 01:31:34.93    & +29:00:31.9 &  2013-10-10 \\ 
M33 21 & 01:31:27.21    & +30:28:51.0 &  2013-10-09 \\
M33 22 & 01:34:42.51    & +31:12:55.7 &  2013-09-28\\
M33 17 & 01:33:45.76    & +32:07:20.8 &  2013-10-13\\

\end{tabular}
\end{table*}


\begin{table*}
\caption{WHT spectroscopy.}             
\label{table:2}      
\centering                          
\begin{tabular}{c c c c c}        
\hline\hline                 
Object & R.A. J2000 & DEC J2000 & $\mathrm{m_{5007}}$& Exp. time (s) \\    
\hline                        

  GCM 1 & 01:34:48.86 & +31:05:14.8 & 20.9  & 3400 \\      
  GCM 2 & 01:33:45.20 & +30:21:22.0 & 22.7  & 3600 \\
  GCM 3 & 01:33:52.30 & +30:21:12.0 & 23.3  & 3600 \\

\hline                                   
\end{tabular}
\end{table*}

\section{Data reduction}

Data were debiased and flat fielded using the ING WFC data reduction pipeline \citep{Irwin2001}.
At each telescope position, images were aligned to one of the [OIII] frames, correcting for geometrical distortions. Then all images in the same filter were averaged. These steps were performed using routines available within \textsc {IRAF{\footnote{\textsc{IRAF} is distributed by the National Optical Astronomy Observatories, which is operated by the Association of Universities for Research in Astronomy, Inc. (AURA) under cooperative agreement with the National Science Foundations. }}}.
Astrometric solutions were computed using the tasks SCAMP and MISSFITS (\citealt{Bertin2006}, \citealt{Marmo2008}) and the SDSS-R9 catalogue \citep{Ahn2012}.
The accuracy of astrometry is $\sim$0.1$"$.
 
Nights were photometric and instrumental magnitudes for the broadband filters were calibrated by matching stars in the observed fields with the SDSS-R9 catalogue, taking advantage of the large number of stars available. For the narrowband H$\alpha$ images, we used the same zero point as that derived for the r' band, as shown in \citet{Drew2005}. The emission-line fluxes in the narrowband filter [OIII] were calibrated using the magnitudes of 31 PNe in \citet{Ciardullo2004}, which are included in our fields 16, 21, and 22. The [OIII]$\lambda5007\AA$ magnitudes ($\mathrm{m_{5007}}$) are related to the monochromatic flux by
\begin{equation}\label{eq:magnitud}
\mathrm{m_{5007}=-2.5\cdot log_{10}\cdot F_{5007}- 13.74}
,\end{equation}
where F$_{5007}$ is expressed in $\mathrm{ergs\cdot cm^{-2}\cdot s^{-1}}$ \citep{Jacoby1989}.\\ 
The photometric solution was then extended to all observed fields using the overlapping areas in different pointings. The
PNe from \citet{Ciardullo2004} used to calibrate our [OIII] images and their corresponding identification number and m$_{5007}$ are shown in Table~\ref{table:oiiicalibration}. Also shown are the corresponding identification numbers in \citet{Magrini2000} and the final magnitudes obtained for our data after the calibration of the fields.


\begin{table*}
\caption{Known PNe used for the [OIII] calibration.}             
\label{table:oiiicalibration}      
\centering                          
\begin{tabular}{c c c c c c}        
\hline\hline                 

 ID(a)      & ID(b)  &   R.A. J2000 &    DEC J2000  &   $\mathrm{m_{5007}}$(a) &  $\mathrm{m_{5007}}$(c) \\   
\hline    

  75   &    67   &    01:33:52.64 &    +30:16:54.2 &    21.05 &     21.06 \\
  60   &    59   &    01:33:44.38 &    +30:20:23.7 &    21.15 &     21.22 \\
  66   &    61   &    01:33:46.73 &    +30:17:33.5 &    21.15 &     21.11 \\
  71   &    63   &    01:33:50.02 &    +30:14:25.2 &    21.25 &     21.23 \\
  11   &    13   &    01:32:55.04 &    +30:09:52.9 &    21.34 &     21.34 \\
  152  &    -    &    01:35:13.63 &    +31:00:48.7 &    21.61 &     21.69 \\
  114  &    89   &    01:34:15.72 &    +31:08:12.5 &    21.72 &     21.79 \\
  117  &    97   &    01:34:18.54 &    +30:58:30.5 &    21.90 &     21.97 \\
  12   &    -    &    01:32:55.11 &    +30:14:01.5 &    22.23 &     22.28 \\
  134  &    107  &    01:34:31.52 &    +31:06:51.3 &    22.33 &     22.32 \\
  133  &    108  &    01:34:31.49 &    +31:05:24.0 &    22.37 &     22.37 \\
  141  &    121  &    01:34:41.96 &    +30:56:49.7 &    22.41 &     22.39 \\
  142  &    120  &    01:34:43.57 &    +31:06:10.7 &    22.51 &     22.52 \\
  147  &    -    &    01:34:58.18 &    +31:06:47.5 &    22.51 &     22.27 \\
  148  &    -    &    01:35:04.92 &    +30:58:42.1 &    22.54 &     22.51 \\
  34   &    33   &    01:33:22.85 &    +30:13:41.0 &    22.59 &     22.49 \\
  6    &    4    &    01:32:42.72 &    +30:12:25.5 &    22.63 &     22.65 \\
  31   &    30   &    01:33:21.16 &    +31:06:44.0 &    22.83 &     22.90 \\
  14   &    15   &    01:32:59.40 &    +30:10:24.0 &    22.84 &     22.82 \\
  104  &    86   &    01:34:11.74 &    +31:07:31.4 &    22.84 &     22.84 \\
  28   &    27   &    01:33:18.48 &    +30:12:38.8 &    23.23 &     23.25 \\
  15   &    16   &    01:33:01.25 &    +30:15:31.1 &    23.42 &     23.50 \\
  1    &    -    &    01:32:09.04 &    +30:22:05.7 &    23.51 &     23.49 \\
  100  &    81   &    01:34:06.76 &    +31:00:29.3 &    23.66 &     23.62 \\
  22   &    19   &    01:33:08.99 &    +30:13:58.6 &    23.78 &     23.41 \\
  72   &    64   &    01:33:50.82 &    +30:18:43.8 &    23.90 &     23.90 \\
  80   &    71   &    01:33:55.58 &    +30:16:02.9 &    23.94 &     23.81 \\
  122  &    100  &    01:34:22.96 &    +30:59:32.9 &    24.07 &     24.16 \\
  129  &    102  &    01:34:25.83 &    +31:07:44.5 &    24.28 &     24.45 \\
  58   &    57   &    01:33:43.48 &    +30:59:41.1 &    24.42 &     24.18 \\
  25   &    25   &    01:33:15.50 &    +30:17:53.7 &    24.51 &     24.42 \\

\hline                                   
\end{tabular}

\begin{flushleft}
(a) \citet{Ciardullo2004};
(b) \citet{Magrini2000};
(c) This work.
\end{flushleft}

\end{table*}

Photometric errors derived internally from DAOPHOT, photon statistics, and background uncertainties are small 
and generally lower than the typical systematic errors. 
We have to account for the uncertainty from the calibration sources in \citet{Ciardullo2004} and the dispersion term for each calculated zero point, including the systematic error of their propagation. Adding all of the sources of uncertainty through least squares, an error up to $\sim$0.1 mag for m[OIII]$\le$24.5 was estimated, increasing to 0.4 mag for m[OIII]$\le$25.5. These errors are larger for faintest sources.

Spectra were reduced (debiased, overscan corrected, trimmed, flat fielded, combined, and finally sky subtracted and extracted) using the packages CCDRED and TWODSPEC in IRAF. Before subtracting the sky background, cosmic rays were eliminated using the task \textit{lacos spec} \citep{Lacos2001}. The flux calibration was carried out using the standard star Feige 15 \citep{Stone1977}. 
More details are given in section 5.2.


\section{Data analysis}
The PN search was made in two different ways: by visual inspection of the combined set of narrow and broadband images and via the analysis of colour-colour diagrams from automated photometric measurements. 

As mentioned before, 31 of the known PNe in M33 are located in a region of about 0.4 square degrees close to the bright disc of the galaxy, which is a zone that was partially covered by three of our mapped regions. Those PNe were used as a reference sample for the visual and photometric searches.

\subsection{Visual search}

The g' images were used as the reference continuum for the [OIII] frames and the r' images were used as reference continuum for H$\alpha$.
Various display colours were assigned to images in different filters to create colour-composite images in which we searched for sources with  H$\alpha$ and [OIII] excesses. Specifically, red was assigned  to the wide filter g' or r', and green to the narrow filter [OIII] or H$\alpha$, which makes point-source emission-line objects appear as outstanding green dots in both the [OIII]/g' and H$\alpha$/r' colour-composite images, whereas normal stars appear with a nearly homogeneous orange colour.

In this way, we identified 8 PN candidates, showing both [OIII] and H$\alpha$ excesses, and  2 additional objects, showing only an [OIII] excess. All these candidates are located at distances smaller than $\sim$9 kpc from the centre of M33, i.e. inside the bright optical disc of the galaxy. 
At distances larger than 9 kpc, no objects with both an [OIII] and H$\alpha$ emission were found. However, 14 additional sources were detected in those external areas with just an [OIII] excess and lower or negligible emission in the other three filters.

\subsection{Colour-colour diagram}

Sources in each of the [OIII] images were automatically searched using SExtractor \citep{Bertin1996}, setting the detection threshold to the limit of 2$\sigma$ over the background. This ensures that very faint sources are recovered, but includes a large number of artefacts. Automatic aperture photometry was then performed via DAOPHOT \citep{Stetson1987} with an empirically determined optimum aperture of $\sim$5/2 times the stellar FWHMs.

\begin{figure}[!h]
  \centering
  \includegraphics[width=\hsize] {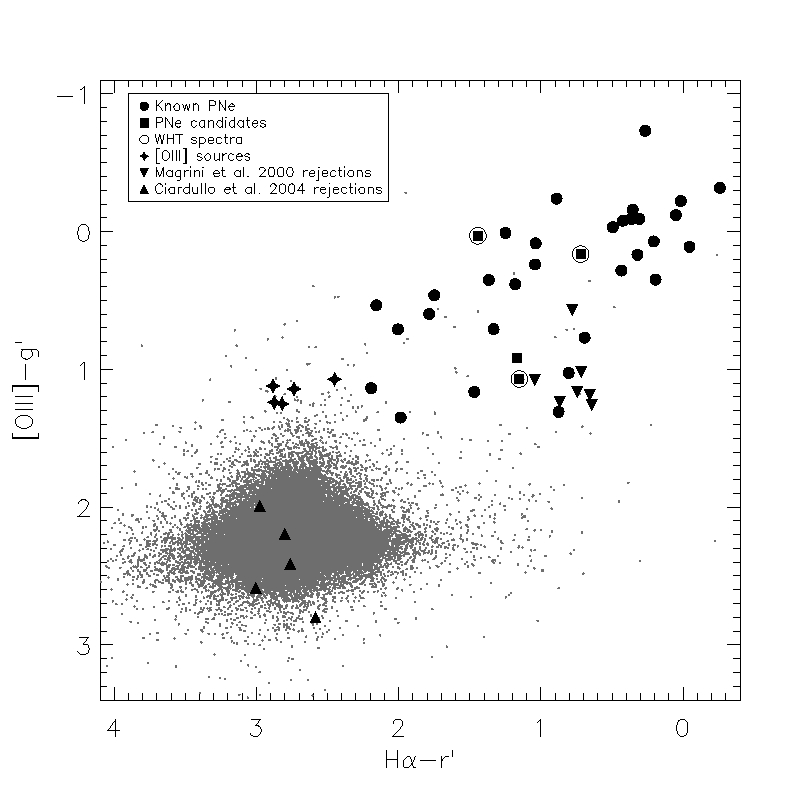}
     \caption{Colour-Colour diagram of the sources in the sampled area of M33.}
     \label{color}
 \end{figure}

Photometry of $\sim$100\,000 stars was obtained and the results were analysed using H$\alpha$-r' versus [OIII] – g' colour-colour diagrams (Fig.~\ref{color}), which are particularly suited for selecting PNe (see e.g. \citealt{Corradi2005}). 


In this kind of diagram, PNe show up as objects with strong [OIII] and H$\alpha$ excess compared to the bulk of normal stars, as indicated by the filled circles in Fig.~\ref{color}, which show the location in the colour-colour diagram of the photometry of the known PNe in the observed fields. All detected sources in this area were  visually inspected in the individual images to reject artefacts or poor-photometry sources. This check was also extended to sources with just [OIII] excess. In this way, 4 new PNe candidates were identified, all of which were independently found by our visual search. In addition, 5 of the 14 sources mentioned above, showing just [OIII] excess,  were also recovered by the photometry study within relatively low errors.

Some candidates found by visual inspection are not recovered by the photometric search. The main reason is that they are exceedingly faint in the broadband filters. In some other cases, they are lost because of confusion in very crowded areas.

\subsection{Contaminants}

The degree of contamination by spurious sources is especially high in the crowded regions of the galaxy and close to spiral arms. This is mostly caused by the presence of giant HII regions and the associated high stellar background. 
Because of the low detection threshold adopted, spurious PN candidates were also found in less crowded regions, mainly at the faintest magnitudes, owing to large photometric errors. Detector cosmetics, cosmic rays, and transient effects such as asteroids or satellite tracks also contribute to contamination.

In order to clean the colour-colour diagram from contaminants, we used the following additional information.
At the distance of M33 (840 kpc, \citealt{Freedman2001}), we do not expect to resolve any PN \citep{Magrini2000}. Therefore every extended source is not considered a PN.
Besides, even though PNe on the Milky Way show a significant emission in the IR, it is not expected that PNe at the distance of M33 would be detected by standard all-sky surveys. Every source with an IR  counterpart in the WISE archive is therefore rejected as a PN. Symbiotic stars can mimic PNe in this survey. Indeed, \citet{Mikolajewska2017} discovered 12 symbiotic stars in M33. 

 Only one of these stars, M33 SySt-11, is associated with a faint WISE source, which is only detected in the shorter w1 and w2 bands.
 The sources showing IR emission (Table~\ref{table_PNe_candidates_LM}) are therefore unlikely to be symbiotic stars.

Even after applying these criteria, some dubious sources remain, and we have visually inspected and searched in catalogues and the literature, all candidates that fall in the selected regions of the colour-colour diagram.

\section{Results}

\subsection{New and rejected PN candidates}

Eight PN candidates were identified by combining the photometric and visual search. Their coordinates, deprojected distances from the centre of the galaxy, and [OIII]$\lambda$5007$\AA$-to-H$\alpha$+[NII] line flux ratios $\mathrm{R}$ are shown in Table \ref{table_PNe_candidates}. To compute the deprojected galactocentric distances, the galaxy distance of 840 kpc \citep{Freedman2001}, an intermediate inclination of 53$^{\circ}$, and the position angle of 22$^{\circ}$ are assumed \citep{Hyperleda2014}.
All candidates have an $\mathrm{R}$ ratio  between the limits $\mathrm{0.3<R<3}$ established by \citet{Ciardullo2002} for PNe within 3 magnitudes from the cut-off of the PN luminosity function. We note that the value of this cut to exclude HII regions decreases at fainter magnitudes \citep{Merrett2006}.

It is important to emphasize that every previously known PN located in our fields were recovered by our search. However, based on our measured colours, we did not confirm some of the candidates in \citet{Ciardullo2004}. These are listed in Table \ref{table_miss}.
Five of these six rejections are shown in the colour-colour diagram of Fig.~\ref{color} as filled upward triangles.
The sixth unconfirmed candidate shows big photometric errors and is rejected as a possible PN after visual inspection.
For the 31 known PNe adopted as the reference sample, the H$\alpha$+[NII] measured fluxes (not shown here) are brighter than in the \citet{Ciardullo2004} results. As explained in \citet{Ciardullo2004}, when compared with \citet{Magrini2000}, this is partly due to the differing contributions of [NII] within the bandpasses of the filters. 

\begin{figure}[h]
  \centering
  \includegraphics[width=\hsize]{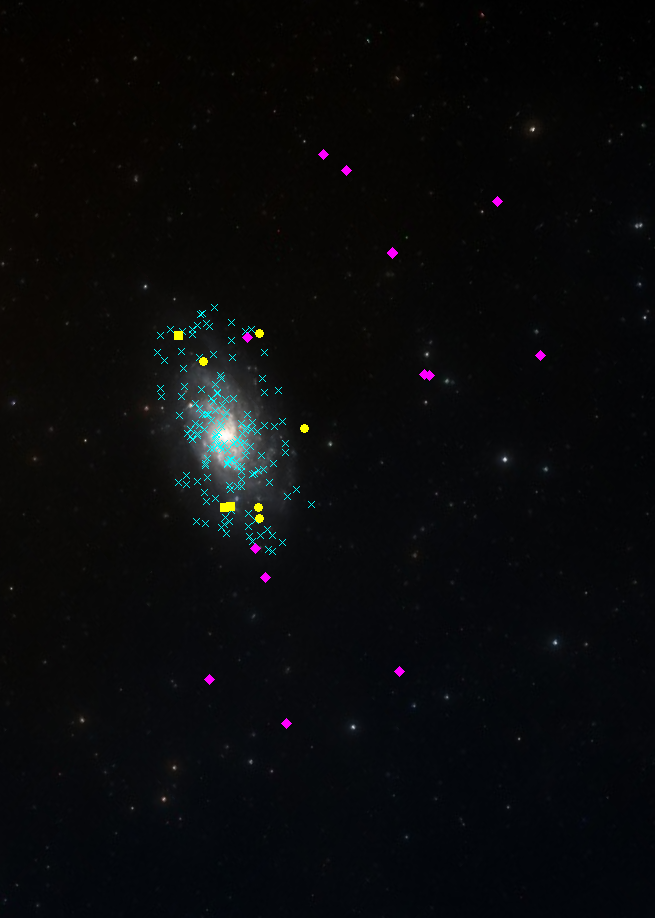}
     \caption{Location of the discovered objects. The yellow filled circles indicate PNe candidates. The yellow filled squares indicate the confirmed PNe and magenta rhombi are candidates with an [OIII] excess. Previous known PNe are shown with cyan small crosses. The total field of view of the image is 3$^{\circ}$ x 4 $^{\circ}$. North is up, east at left.
           }
        \label{location}
\end{figure}

Several candidates with apparently good colour data were also rejected after visual inspection.  Seven of these candidates are objects previously found by \citet{Magrini2000} (filled downward triangles in Fig. 2): five were listed as ``possible PNe'' by \citet{Magrini2000}, which had been previously discarded as PNe by \citet{Ciardullo2004}, and two more were catalogued as emission-line objects with non-negligible continuum (Table~\ref{table_PNe_candidates_LM}).
Additional compact sources with only [OIII] excess are listed in Table~\ref{table_O_excess}. Some have an H$\alpha$ and r' counterpart, but show no detectable H$\alpha$ excess. These could be compact galaxies with [OII] $\lambda$3727$\AA$ emission redshifted to $z\sim0.35$ or Ly$\alpha$ emitters at $z\sim3.1$. Another less likely possibility is that these sources are M33 H-deficient PNe. Further studies of these sources are of interest.
The locations of the discovered objects and the previously known PNe are shown in Fig.~\ref{location}. Finding charts for our PNe candidates (Table~\ref{table_PNe_candidates}) are shown in Fig.~\ref{fig:finding}, whereas images of three [OIII] excess sources (Table~\ref{table_O_excess}) are shown in Fig.~\ref{fig:finding[OIII]}. Finally, Fig.~\ref{fig:finding4colours} shows images at the four bands of a PN and an excess [OIII] source.

\newpage
\clearpage


\begin{table*}[!h]
\caption{PNe candidates.}             
\label{table_PNe_candidates}      
\centering                          
\begin{tabular}{c c c c c c c}        
\hline\hline                 
Object & R.A. J2000 & DEC J2000 & d (kpc) & $\mathrm{m_{5007}}$ & F([OIII])/F(H$\alpha$) & Comments  \\    
\hline                        

GCM 1  & 01:34:48.84 & +31:05:14.9 &  6.99   &  20.9 & 2.39  & Confirmed PN$^\ast$, WHT spectra \\

GCM 2  & 01:33:45.18 & +30:21:21.4 &  4.83   &   22.7 &  0.73 & Confirmed PN, WHT spectra\\

GCM 3  & 01:33:52.30 & +30:21:12.2 &  5.06   &   23.3  & 1.57 & Confirmed PN, WHT spectra \\

GCM 4  & 01:33:11.37 & +30:21:12.6 &  4.96   &  24.0   & 0.52 &   Crowded region \\ 

GCM 5  & 01:32:17.29 & +30:41:35.0 &  8.00   &  25.2  & 1.80 & Visual search. Extremely faint. \\ 

GCM 6  & 01:33:11.52 & +31:05:49.2 &  8.83   &   25.4  & 0.72 & Visual search. Extremely faint. \\ 

GCM 7  & 01:34:18.25 & +30:58:27.2 &  4.85   &  25.7     & 0.63  & Visual search. Extremely faint. \\ 

GCM 8 & 01:33:11.22 & +30:18:28.8 &  5.57   &   24.4   & - 
  & Visual search. Crowded region \\ 

\end{tabular}

\begin{flushleft}
$^\ast$PN128, classified as ``possible PN'' in \citet{Magrini2000}
\end{flushleft}

\end{table*}


\begin{table*}[!h]
\caption{PNe misidentified in \citet{Ciardullo2004}. }            
\label{table_miss}      
\centering                          
\begin{tabular}{l c c c}        
\hline\hline                 
Object & R.A. J2000 & DEC J2000  \\    
\hline                        

PN 151 & 01:35:10.50 & +31:05:14.9  \\
PN 144 & 01:34:45.50 & +31:01:11.7  \\
PN 120 & 01:34:20.50 & +31:10:52.0  \\
PN 97  & 01:34:05.54 & +31:12:30.1  \\
PN 61  & 01:33:44.62 & +31:04:03.5  \\
PN 42  & 01:33:27.89 & +31:06:22.3  \\

\end{tabular}
\end{table*}


\begin{table*}[!h]
\caption{Rejected objects classified as ``possible PNe'' by \citet{Magrini2000}.}             
\label{table_PNe_candidates_LM}      
\centering                          
\begin{tabular}{l c c c}        
\hline\hline                 
Object & R.A. J2000 & DEC J2000 & Comments  \\    
\hline                        

 PN23  & 01:33:13.54 & +30:22:36.1  & Extended, IR counterpart (w3, w4) \\
 PN5   & 01:32:44.16 & +30:22:03.9  & IR counterpart (w3, w4) \\
 PN127 & 01:34:47.10 & +30:59:36.2  & Extended, Galaxy 2247 \\
 PN78  & 01:34:02.55 & +30:58:10.5  & Extended, IR counterpart (w3, w4), CO counterpart \\
 PN55  & 01:33:41.77 & +30:08:31.3  & Extended, IR counterpart (w1, w2)\\
 Em24  & 01:33:10.78 & +30:18:08.8  & Globular cluster  \\
 Em59  & 01:34:42.52  & +30:55:44.4 & Extended, globular cluster candidate  \\

\end{tabular}
\end{table*}


\begin{table*}[!h]
\caption{Sources showing a substantial [OIII] excess.}             
\label{table_O_excess}      
\centering                          
\begin{tabular}{l c c c c c}        
\hline\hline                 
Object & R.A. J2000 & DEC J2000 & d (kpc) & $\mathrm{m_{5007}}$ & Comments  \\    
\hline                        
GCM E1 & 01:33:02.82   & +30:03:15.5 &  9.32  & 24.8  & Detected in g' and r'. No H$\alpha$. Visual search. \\

GCM E2 & 01:27:35.41   & +31:00:08.2& 33.60  & 24.1  & Detected in g' and r'. No H$\alpha$. Visual search. \\

GCM E3 & 01:31:27.30   & +31:47:39.5 & 25.24  & 24.8  & Detected in g'. No H$\alpha$, no r'. Visual search. \\ 

GCM E4 & 01:28:26.44   & +31:39:31.8 & 35.93 & 24.7   & No H$\alpha$, no r', no g'. Visual search. \\ 

GCM E5 & 01:29:48.03   & +30:55:06.1 & 22.03  & 24.7  & Detected in g' and r'. No H$\alpha$. Visual search.  \\

GCM E6 & 01:29:54.43   & +30:55:17.2 & 21.54  & 25.4  & Detected in g' and r'. No H$\alpha$. Visual search.   \\

GCM E7 &  01:34:09.30  & +29:37:10.3 & 17.63  & 24.9  &  Detected in g' and r'. No H$\alpha$. Visual search. \\ 

GCM E8 &  01:30:31.82  & +31:26:45.3 & 24.17  & 24.8  & Detected in g'. No H$\alpha$, no r'. Visual search. \\ 

GCM E9  & 01:33:14.80 & +30:10:39.4 &  7.42   &   25.4  & Bright in r', g', H$\alpha$. Visual search. \\

GCM E10 & 01:33:25.78   & +31:04:51.6 &  7.86  & 24.0  & Bright in r', g', H$\alpha$.\\

GCM E11{$^\ast$} & 01:32:38.31 & +29:26:02.5 & 18.85  & 24.6  & Bright in r', g', H$\alpha$. \\ 

GCM E12  & 01:31:55.81 & +31:51:51.8 & 24.66  & 24.3  & Bright r', g', H$\alpha$. \\

GCM E13 & 01:30:32.15 & +31:26:18.7 & 24.05  & 24.9  & Bright r', g', H$\alpha$.\\

GCM E14 & 01:30:24.77 & +29:39:18.6 &   19.32  & 23.8    & Bright r', g', H$\alpha$. \\

\end{tabular}

\begin{flushleft}
{$^\ast$}J013238.33+292602.57, catalogued as quasar at z$\sim$4.45 by \citet{Wang2016}. It should be revised with the detected [OIII] excess.
\end{flushleft}

\end{table*}

\clearpage
\newpage


\subsection{Spectroscopic confirmation of PN candidates}

\begin{figure*}[h]
   \centering
   \includegraphics[width=15cm, height=11cm]{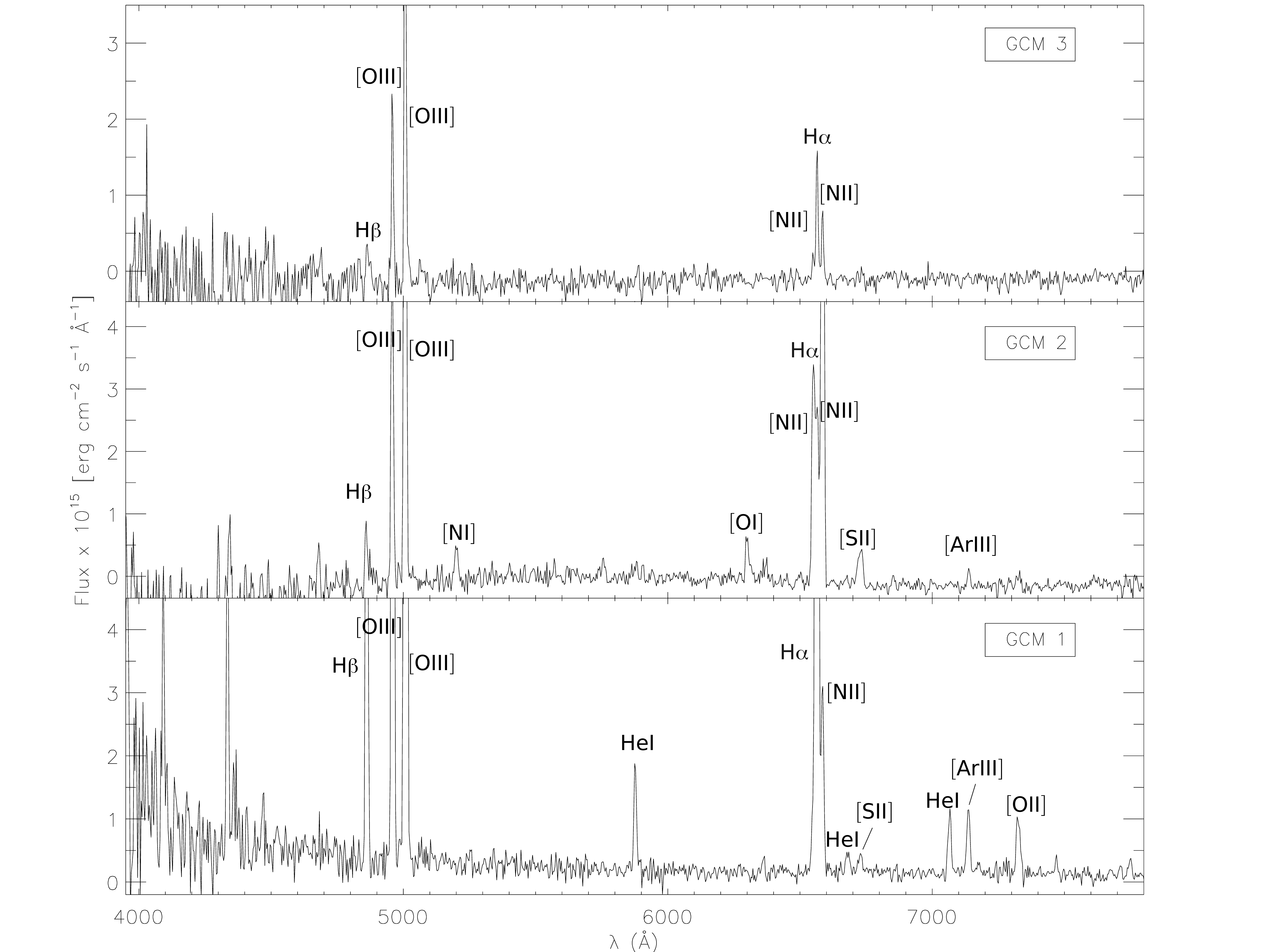}
      \caption{WHT spectra of three new PNe. We note the extremely bright [NII] lines in GCM2.}
         \label{espectros}
   \end{figure*}

\begin{figure*}[!htbp]
  \centering
  \subfigure[GCM1]{\includegraphics[width=4.0cm, height=4.0cm]{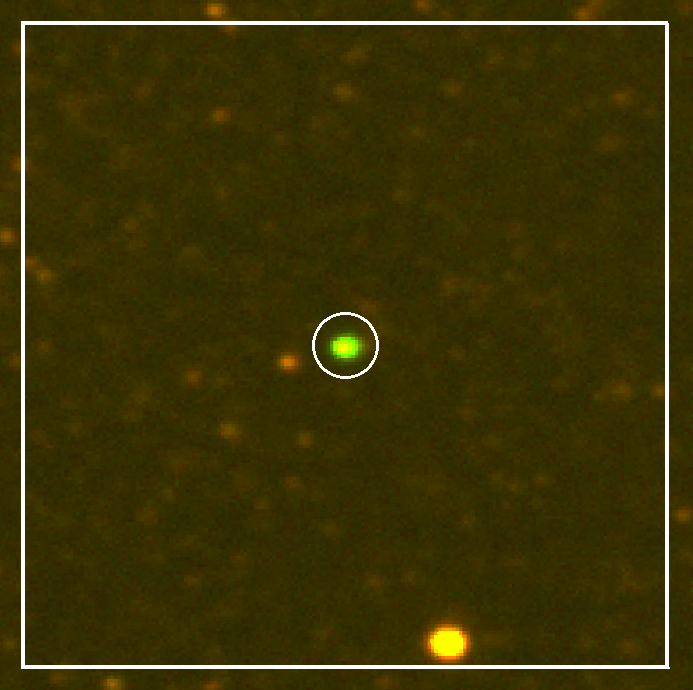}}\label{finding1}
  \subfigure[GCM2]{\includegraphics[width=4.0cm, height=4.0cm]{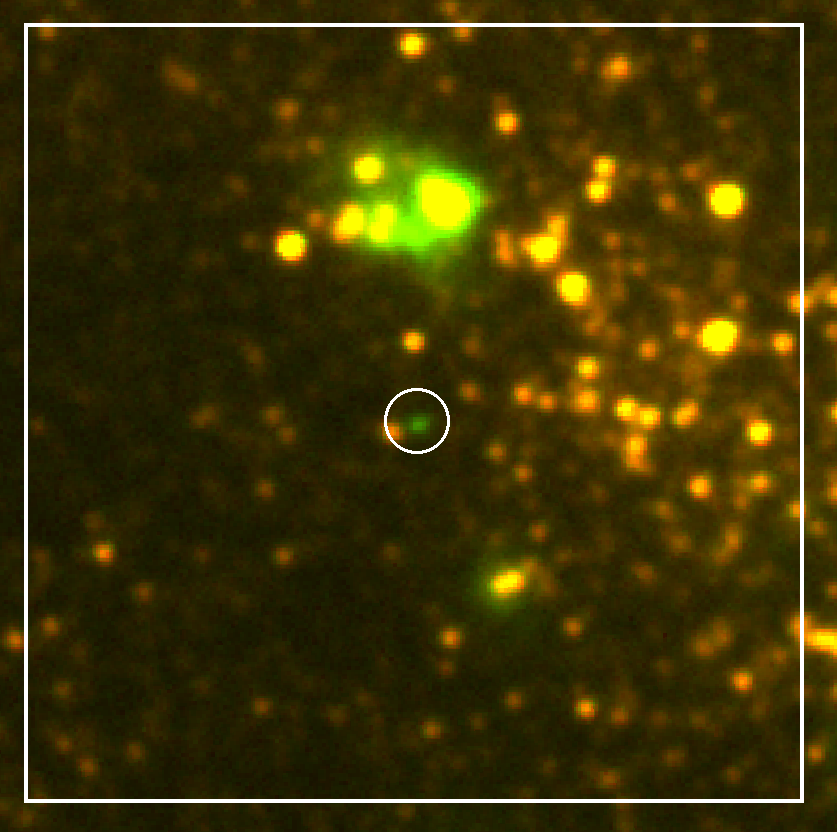}}\label{finding2}
  \subfigure[GCM3]{\includegraphics[width=4.0cm, height=4.0cm]{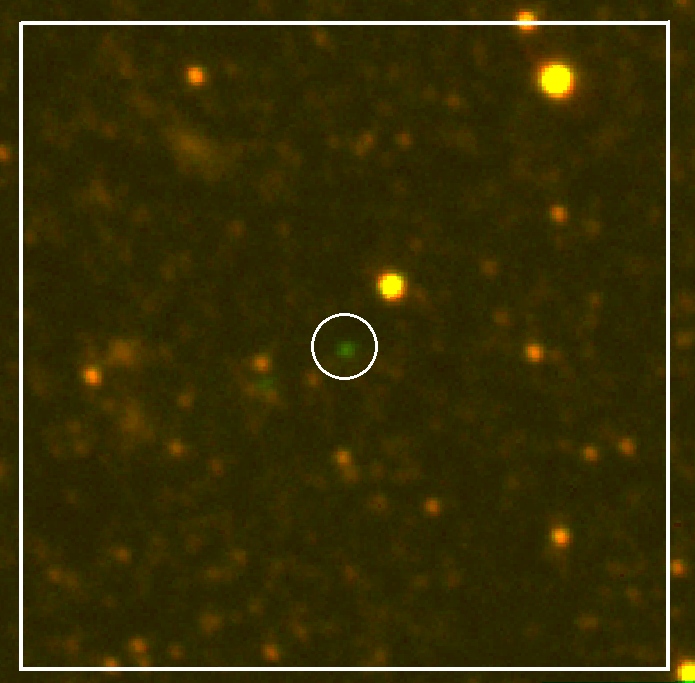}}\label{finding3}
  \subfigure[GCM4]{\includegraphics[width=4.0cm, height=4.0cm]{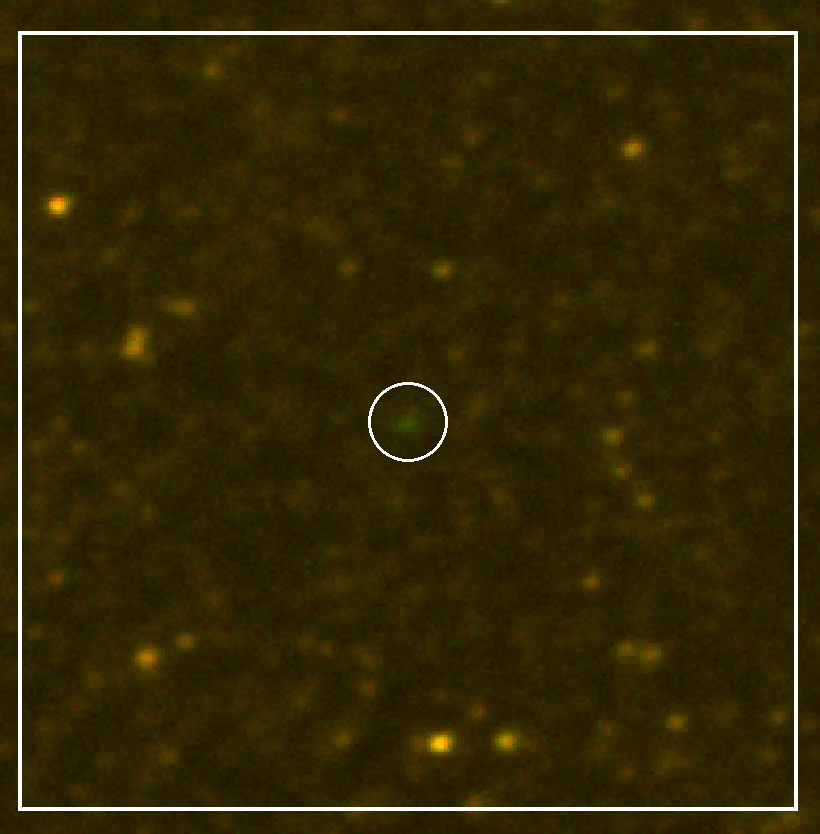}}\label{finding4}
  \subfigure[GCM5]{\includegraphics[width=4.0cm, height=4.0cm]{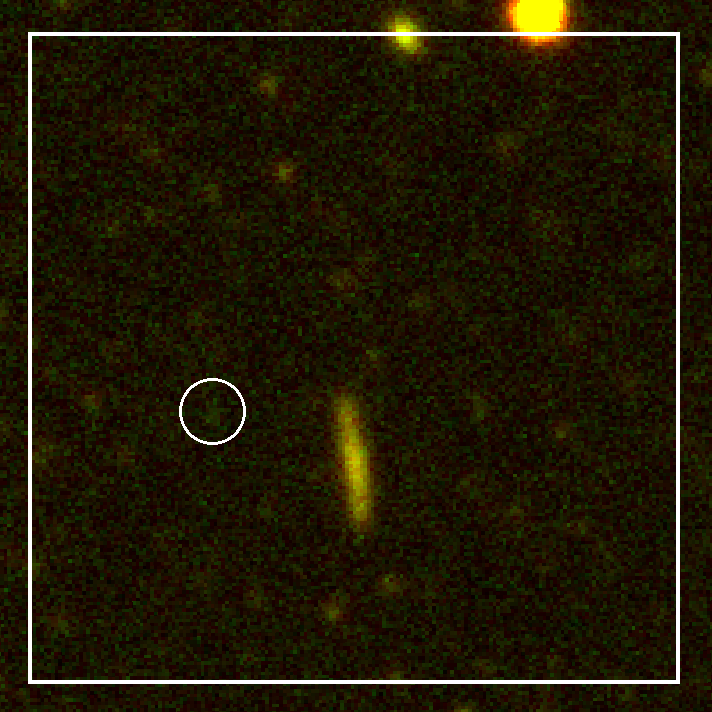}}\label{finding5}
  \subfigure[GCM6]{\includegraphics[width=4.0cm, height=4.0cm]{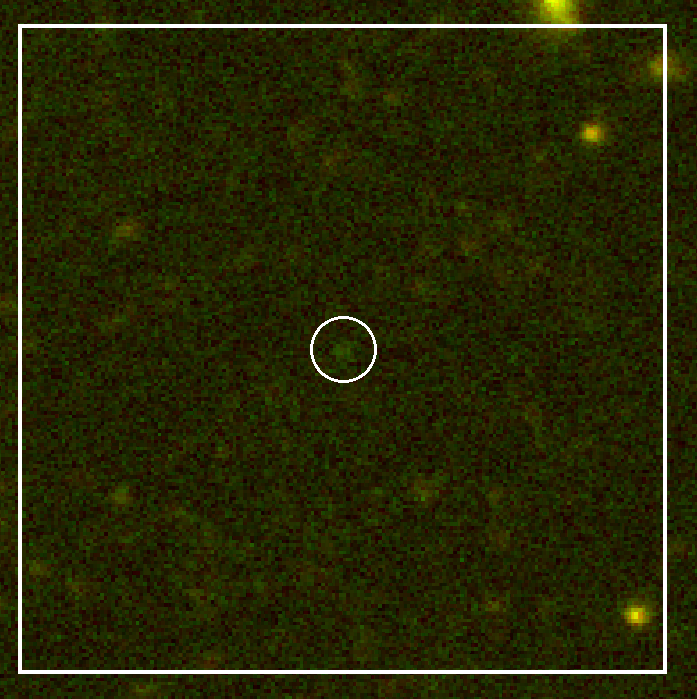}}\label{finding6}
  \subfigure[GCM7]{\includegraphics[width=4.0cm, height=4.0cm]{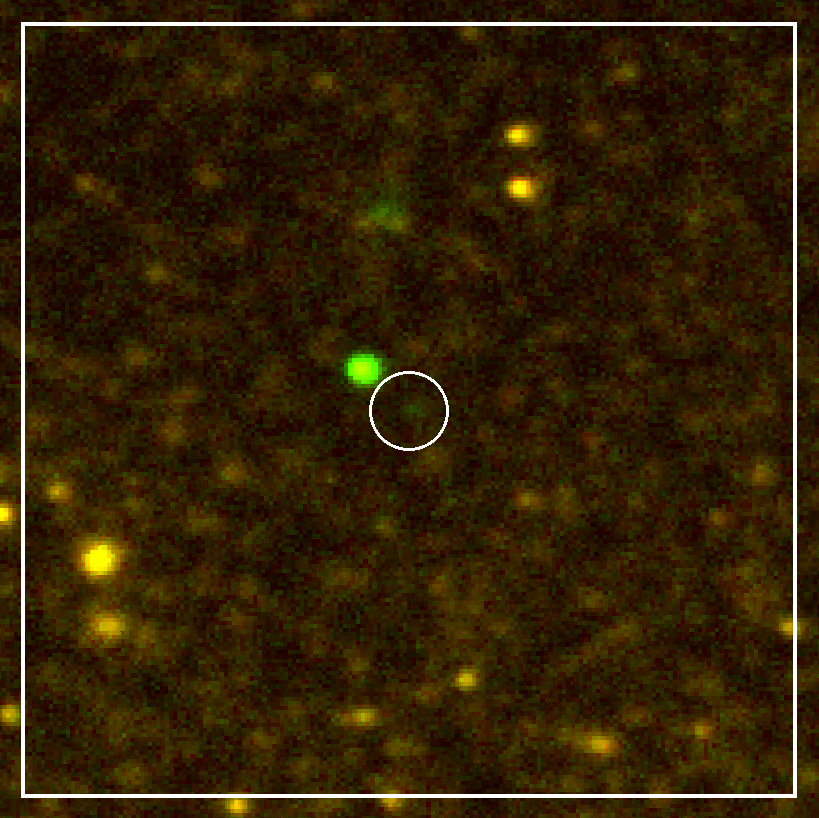}}\label{finding7}
  \subfigure[GCM8]{\includegraphics[width=4.0cm, height=4.0cm]{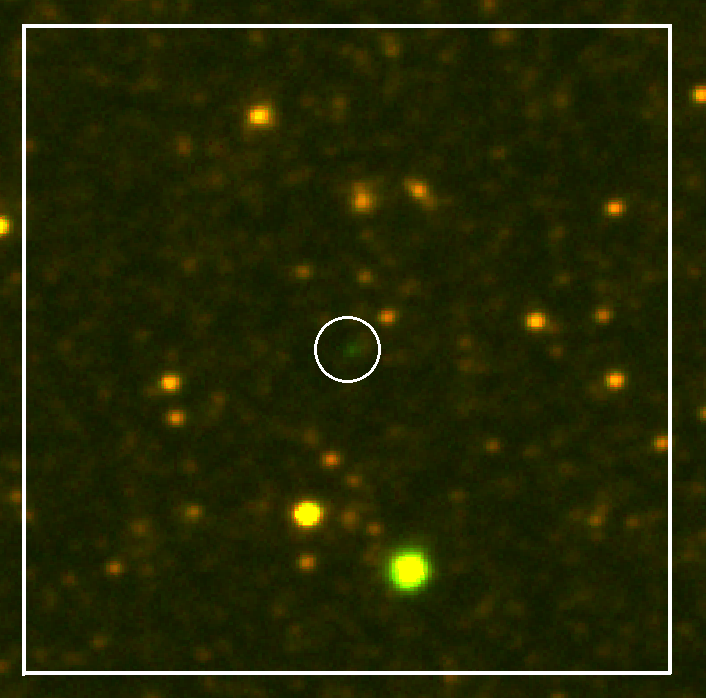}}\label{finding8}
  \caption{Finding charts for the PNe candidates (green: [OIII], red: g'). North is up, east at left. Field size is 1’ x 1’.}\label{fig:finding} 
 \end{figure*}

\begin{figure*}[!htbp]
  \centering
  \subfigure[GCM E2]{\includegraphics[width=4.0cm, height=4.0cm]{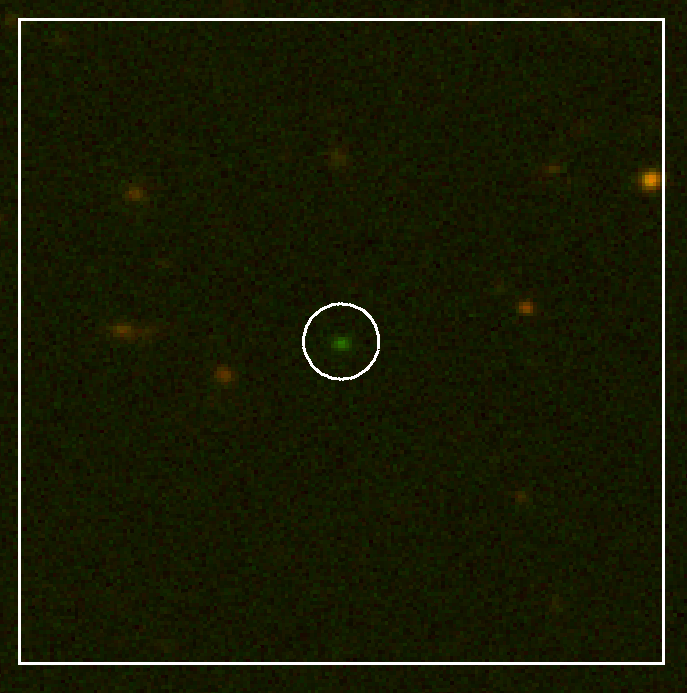}}\label{findingE2}
  \subfigure[GCM E5]{\includegraphics[width=4.0cm, height=4.0cm]{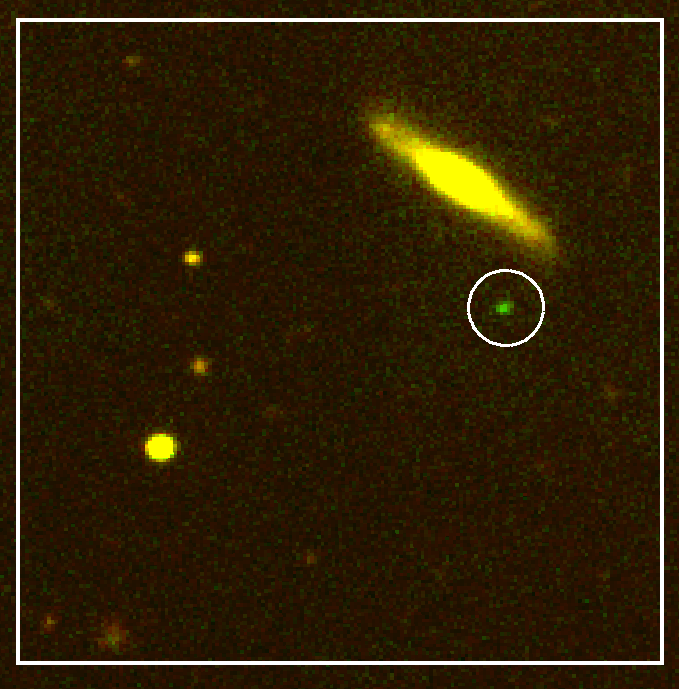}}\label{findingE5}
  \subfigure[GCM E6]{\includegraphics[width=4.0cm, height=4.0cm]{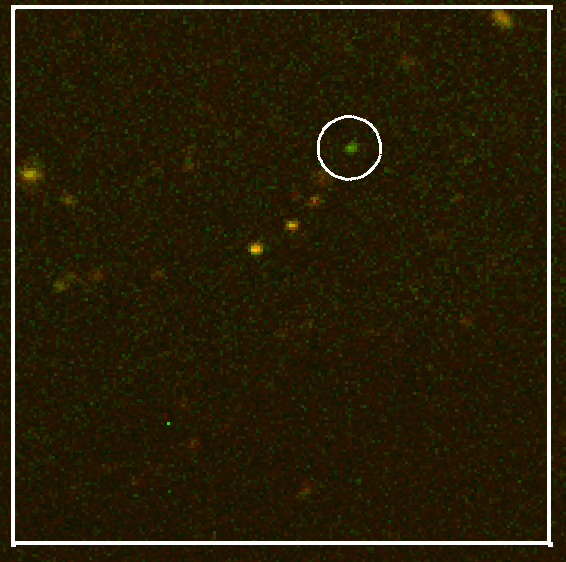}}\label{findingE6}
   \caption{As in Fig. 5 for three sources with [OIII] excess.}\label{fig:finding[OIII]} 
 \end{figure*}

\begin{figure*}[!htbp]
  \centering
  \subfigure[GCM2 (PN).]{\includegraphics[width=6.5cm, height=6.5cm]{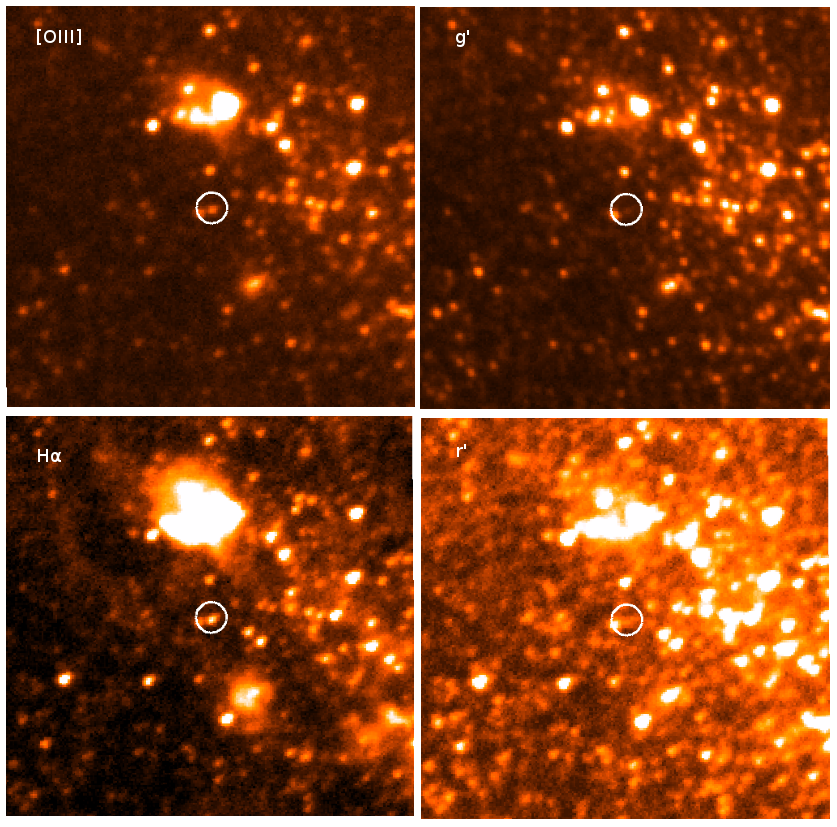}}\label{finding4coloursPNe}
  \subfigure[GCM E2 ($\mathrm{[OIII]}$ excess source).]{\includegraphics[width=6.5cm, height=6.5cm]{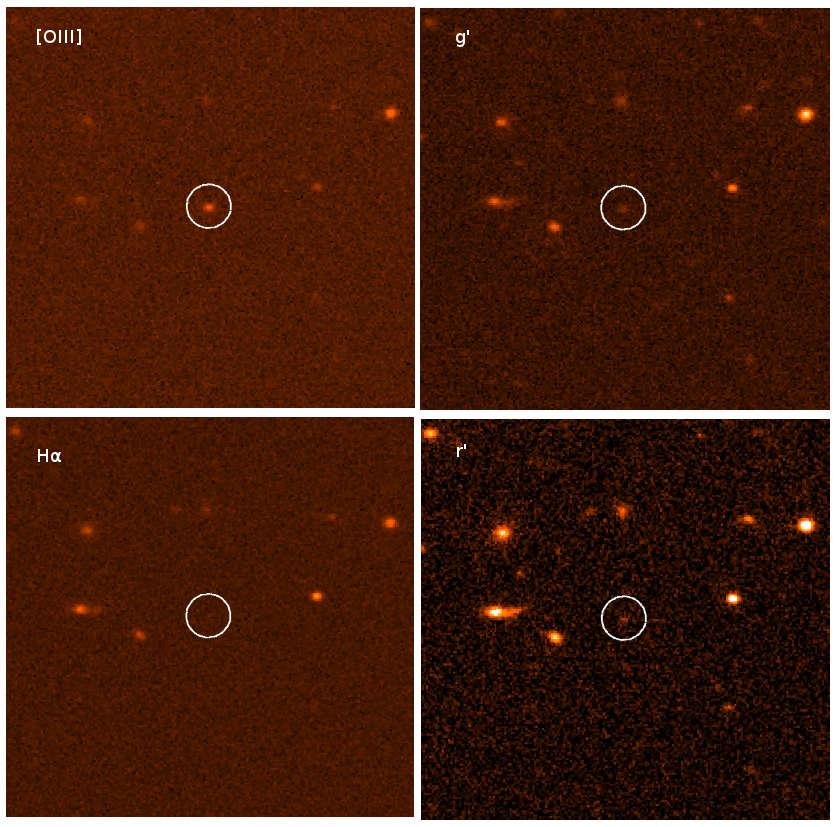}}\label{finding4coloursE2}
  \caption{Two representative sources imaged at the four filters. North is up, east at the left. Field size is 1’ x 1’.}\label{fig:finding4colours} 
 \end{figure*}

The three brightest candidates were spectroscopically confirmed using ACAM at the WHT. Spectra are shown in Fig.~\ref{espectros}. 
Relative flux calibration was carried out using the standard star Feige 15 \citep{Stone1977}, observed on the night of December 27,$\mathrm{^{}}$ 2014.\\ 
Emission-line fluxes were measured using the package SPLOT of IRAF. Errors in the fluxes were estimated adding the measured statistical errors quadratically to a constant 5$\%$ term taken as representative of systematic errors in the flux calibration. Observed line fluxes were corrected for the effect of the interstellar extinction using the extinction law of \citet{Cardelli1989} with R$_{\nu}$=3.1. 
The logarithmic nebular extinction c(H$\beta$) was derived with the observed-to-theoretical Balmer ratio H$\alpha$/H($\beta$) from \citet{Osterbrock2006}. 
 Line identification, rest-frame wavelengths, and the measured and extinction-corrected flux intensities normalized to H$\beta$=100 are presented in Table~\ref{table_spectra_measurements}. 


Fig.~\ref{espectros} shows that all bright emission lines typical of PNe are detected in our low resolution spectra that have [OIII] much brighter than H$\alpha$. Also, in the log(H$\alpha$/[SII]$\lambda\lambda$6717+6731) versus log(H$\alpha$/[NII]$\lambda$6583) diagnostic diagram \citep{Viironen2007} GCM 1 and 2 can be located near to the region populated by PNe. We note that GCM 1 was previously identified by \citet{Magrini2000} as a possible PN candidate, but rejected by \citet{Ciardullo2004} because of its bright emission in their wide filters; in addition,  GCM 2 shows a remarkably high [NII] $\lambda$6548 and $\lambda$6584 flux relative to H$\alpha$, suggesting it is a type I PN.

Given the faintness of the PNe detected lines and limited spectral resolution, no reliable determination of electron temperature ($\mathrm{T_{e}}$) and electron density ($\mathrm{N_{e}}$) is possible.

        
\begin{table*}[h]
\caption{Observed and dereddened fluxes.}             
\label{table_spectra_measurements}      
\centering                          
\begin{tabular}{c c l l r@{\hspace{2pt}} c@{\hspace{1pt}} l r@{\hspace{2pt}} c@{\hspace{1pt}} l}        

\hline\hline                 
Object & c & Ion & $\mathrm{\lambda_{rest}(\AA)}$ &  &  $\mathrm{F_{\lambda}}$ & & & $\mathrm{I_{\lambda}}$ & \\    
\hline
   
 GCM 1 & 0.08$\pm$0.02   &  HI(H$\beta$)    &       4861.3     &    100  & $\pm$     & 6      &   100  & $\pm$   &  6        \\
       &                 &  [OIII]          &       4958.9     &    330  & $\pm$     & 17     &   328  & $\pm$   &  17       \\
       &                 &  [OIII]          &       5006.8     &    1000 & $\pm$     & 50     &   992  & $\pm$   &  50       \\
       &                 &  HeI             &       5875.6     &    17   & $\pm$     & 2      &   16   & $\pm$   &  2       \\
       &                 &  [NII]           &       6548.0     &    13   & $\pm$     & 2      &   12   & $\pm$   &  2      \\
       &                 &  HI(H$\alpha$)   &       6562.8     &    302  & $\pm$     & 16     &   286  & $\pm$   &  16       \\
       &                 &  [NII]           &       6583.4     &    28   & $\pm$     & 2      &   26   & $\pm$   &  2       \\
       &                 &  HeI             &       6678.2     &    4.2  & $\pm$     & 1.2    &   4.0  & $\pm$   &  1.1     \\
       &                 & [SII]            &      6717+6731   &    3.8  & $\pm$     & 1.4    &   3.6  & $\pm$  &  1.3     \\
       &                 & HeI              &       7065.3     &    11   & $\pm$     & 2      &   10   & $\pm$  &  2      \\
       &                 & [ArIII]          &       7135.8     &    12   & $\pm$     & 1      &   11   & $\pm$  &  1       \\
       &                 & [OII]            &       7323.0     &    13   & $\pm$     & 2      &   12  & $\pm$  &  2      \\
       &                 & [ArIII]          &       7751.1     &    3.4  & $\pm$     & 1.4    &   3.2  & $\pm$  &  1.3     \\

\hline
  
GCM 2  & 0.17$\pm$0.10    & HI(H$\beta$)     &      4861.3     &       100     & $\pm$  &  12     &    100     &   $\pm$ &  12     \\
       &                  & [OIII]           &      4958.9     &       581     & $\pm$  &  40     &    575     &   $\pm$ &  35     \\
       &                  & [OIII]           &      5006.8     &       1613    & $\pm$  &  90     &    1589    &   $\pm$ &  90     \\
       &                  & [NI]             &      5200.3      &       84     & $\pm$  &  11     &    81      &   $\pm$ &  10   \\
       &                  & [NII]            &      5755.0      &       41     & $\pm$  &  14     &    38      &   $\pm$ &  13   \\
       &                  & HeI              &      5875.6     &         27    & $\pm$  &  8      &    25      &   $\pm$ &  8    \\
       &                  & [OI]             &      6300.3      &       102    & $\pm$  &  13     &    93      &   $\pm$ &  13   \\
       &                  & [NII]            &      6548.0     &       496     & $\pm$  &  40     &    442     &   $\pm$ &  40     \\
       &                  & HI(H$\alpha$)    &      6562.8     &       321     & $\pm$  &  30     &    286     &   $\pm$ &  30     \\
       &                  & [NII]            &      6583.4     &       1342    & $\pm$  &  70     &    1194    &   $\pm$ &  100   \\
       &                  & [SII]            &     6717+6731    &       130    & $\pm$  &  16     &    115     &   $\pm$ &  15    \\
       &                  & [ArIII]          &      7135.8     &       31      & $\pm$  &  9      &    26      &   $\pm$ &  8    \\
       &                  &  [OII]           &      7323.0      &       37     & $\pm$  &  11     &    32      &   $\pm$ &  10   \\
       
\hline       

GMC3 &  0.00 $\pm$0.24    & HI(H$\beta$)  & 4861.3  &   100     & $\pm$  & 21    &  100     & $\pm$  & 21   \\  
     &     & [OIII]        & 4958.9  &   363     & $\pm$  & 30    &  363     & $\pm$  & 30 \\
     &     & [OIII]        & 5006.8  &   992     & $\pm$  & 60    & 992     & $\pm$  & 60 \\
     &     & [NII]         & 6548.0  &   51      & $\pm$  & 11  &   51      & $\pm$  & 11  \\
     &     & HI(H$\alpha$) & 6562.8  &   265     & $\pm$  & 23    &   265     & $\pm$  & 23 \\
     &     & [NII]         & 6583.4  &   145     & $\pm$  & 16  & 145     & $\pm$  & 16   \\
 
   \end{tabular}
 \end{table*}

\subsection{Completeness limit}

The completeness limit of our search was estimated by adding artificial stars with various magnitudes in both narrowband and continuum images and then computing the recovery rate of such artificial objects (\citealt{Minniti1997}, \citealt{Magrini2002}, \citealt{Magrini2003}, \citealt{Corradi2005}). A recovery rate of ~50$\%$ defines incompleteness.

For [OIII] and H$\alpha$, the completeness limit in our observed regions is between 25.5 and 26.0  mag. In the $\sim$0.4 square degree area closer to the bright disc of the galaxy, this completeness limit typically increases by 0.5 magnitudes, owing to the higher background and crowdedness.
For g' and r', the completeness limit is at 24.5 mag, becoming almost one magnitude brighter, $\sim$23.5, in that 0.4 square degree area closer to the bright disc of M33.

\subsection{PN population size and luminosity of the external regions of M33}

The luminosity of a stellar population that produces PNe can be determined using the $\alpha$ ratio, i.e. the luminosity-specific PN number density \citep{Jacoby1980}. This quantity links the PNe population size with the bolometric luminosity of the parent stellar population. 
\citet{Buzzoni2006} discusses the method in detail and provides recipes to determine $\alpha$ as a function of the age and metallicity (colour) of the host galaxy.

Based on the simple stellar population models by \citet{Renzini1986}, the number of PNe related to its progenitor stellar population can be expressed in terms of the PN visibility lifetimes ($\mathrm{\tau_{PN}}$) as follows:
\begin{equation}\label{eq:alpha}
\mathrm{\alpha = \frac{N_{PN}}{L_{TOT}}=B \cdot\tau_{PN}}
,\end{equation}

where B is the specific evolutionary flux ($\mathrm{stars\cdot years^{-1}\cdot {L_{\odot}}^{-1}}$), that is only weakly dependent on the metallicity (except at very low metallicities).

Given that no PN candidates are found outside the bright M33 disc down to the completeness limit of our survey, we can estimate an upper limit to the total stellar luminosity at the outer regions that we have imaged. For that, we adopt a standard shape for the PNLF and a completeness limit of 25.5 mag, which corresponds to 5 magnitudes below the bright cut-off of the PN luminosity function (PNLF) assuming an absolute magnitude cut-off of $\mathrm{M^{*}=-4.47}$ \citep{Ciardullo2002} at the distance of M33 \citep{Ciardullo2004}. Also, we assume that the faintest detectable PNe are about 8 magnitudes below the bright cut-off $\mathrm{m*}$ of the PNLF \citep{Buzzoni2006}.

From previous deep studies of M33 we know that the colour of the external regions is $\mathrm{(B-V)_{outskirts}\sim0.55}$, corresponding to $\mathrm{\log\alpha\approx-6.8}$ \citep{Buzzoni2006}.
With all these figures, we infer an upper limit for the luminosity of the observed 4.5 square degree region of $\mathrm{\sim1.6\cdot10^{7}L_{\odot}}$.

This agrees with a recent deep study of the RGB population in M33, setting a limit to the luminosity of a possible halo to $\mathrm{< 1.6\cdot 10^{7} L_{\odot}}$ \citep{McMonigal2016} but leaving open the alternative existence of a very extended and faint thick disc (\citealt{Barker2011}, \citealt{Cockcroft2013}).
This value is smaller than, but roughly compatible with, the luminosity of the 40 square degrees substructure surrounding M33 ($\mathrm{\sim1.9\cdot10^{7} L_{\odot}}$) measured by \citet{McConnachie2009} based also on the study of their RGB population.
 In the past, several works have claimed the detection of a halo in M33. Studies of RR Lyrae stars suggested the existence of this halo (\citealt{Sarajedini2006}, \citealt{Yang2010}); subsequent works using metallicity indicators have also suggested its existence \citep{Pritzl2011}. Globular clusters  \citep{Chandar2002} and AGB and RGB population studies (e.g \citeauthor{Barker2007a}\citeyear{Barker2007a}, \citeyear{Barker2007b}, \citealt{Ibata2007} or \citealt{Cioni2008}) also claimed tentative detections of the halo. However, the present work and that from \citet{McMonigal2016} have failed to measure any bright stellar component in the external regions of the galaxy, excluding the existence of a massive classical monolithic halo.

\section{Summary and conclusions}

Planetary nebulae are known to be excellent tracers of stellar populations even in very low surface brightness regions. Taking advantage of this property, we have carried out the first deep search for PNe in the outer regions of M33 to gain further insights into the existence of a classical halo or, as in the case of M31, of a rich system of extended substructures possibly related to past interactions between the two galaxies.

The lack of reliable PN candidates in the observed external regions of M33, and down to the completeness limit of our search of 25.5 mag in [OIII] and H$\alpha$, poses an upper limit to the luminosity of any stellar population in this area of $\mathrm{L_{outskirts, PNe} < 1.6\cdot 10^{7}L_{\odot}}$ $\mathrm{(M_{V}\sim-12.45)}$. It can therefore be concluded that there is no evidence of prominent extended structures outside its bright disc, and particularly, of a massive classical halo, in line with results obtained using other stellar tracers.

The difference with M31, which instead displays a very rich circumgalactic environment populated by bright PNe, is therefore substantial, and goes against the idea of a strong tidal interaction in a past encounter of the two galaxies \citep{Patel2017}.

\newpage

\begin{acknowledgements}
   \ We deeply appreciate discussion with Bruce Balick and Karen Kwitter, which led to the development of this work.\\
     We thank Robert Greimel for the first treatment of the images.\\  
    R.G-R thanks José Antonio Acosta Pulido for the encouragement to apply for the observing time of the E2 source and Ismael Pérez Fournón for his help with the analysis of the quasar's spectra.\\
     We thank Laura Magrini, the referee of this paper, for the detailed revision and suggestions that helped to significantly improve the article.\\
     The work was supported by the Spanish project AYA2012-35330.
    
\end{acknowledgements}

%
%

\newpage
\clearpage


\newpage
\clearpage

\bibliographystyle{aa} 
\bibliography{aa.bib} 

\begin{appendix}

\section{Spectra of a type II QSO candidate.}

A 3$\times$2400 sec spectrum of the [OIII] excess source E2 in Table~5 
was taken on August 14 and September 2, 2017 with the same set-up as described in Sect.~2. 
The standard star HD 19445 \citep{OKeandGunn1983} was used for flux calibration. 
The spectrum is shown in Fig.~\ref{fig:qso}.
Despite the faintness of the source, the characteristic lines of a typical active galactic nucleus (AGN) \citep{Hainline2011}, at redshift z$\sim$3.12, are recognized. 
The rest frame and observed wavelength, corresponding ion, and measured flux are tabulated in Table \ref{table:qso}. The FWHM of the critical Ly$\alpha$ and CIV lines are included.
Type II AGNs are usually identified by their narrow emission lines (FWHM $\mathrm{< 2000 Km\cdot s^{-1}}$) and sources as faint as this object are barely detected \citep{Alexandroff2013}. 

\renewcommand{\thefigure}{A\arabic{figure}}
\setcounter{figure}{0}

\begin{figure*}[!h]
\centering
  \includegraphics[width=17cm, height=6cm]{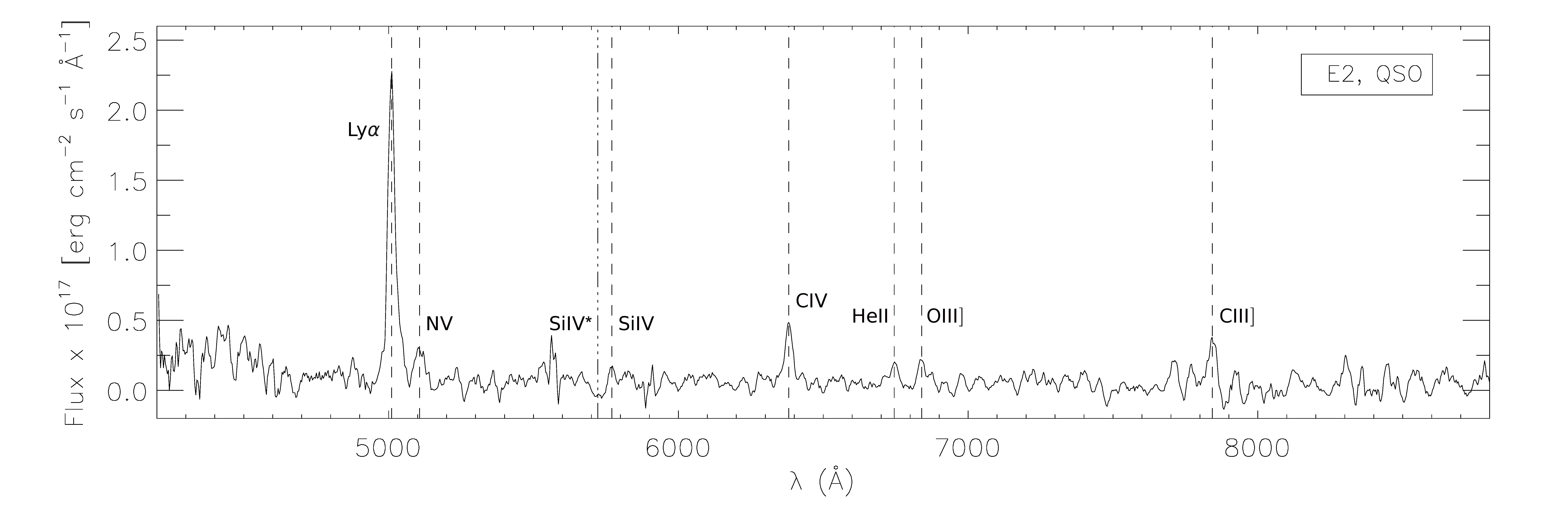}
    \caption{Spectrum of the [OIII] excess source E2, a type II QSO candidate.}
     \label{fig:qso}
\end{figure*}

\renewcommand{\thetable}{A\arabic{table}}
\setcounter{table}{0}

\begin{table*}[!h]
  \caption{Emission and absorption detected features. }            
  \label{table:qso}      
   \centering                          
\begin{tabular}{l c c c D{,}{\hspace{0.3em}\pm\hspace{0.3em}}{-1} } 
\hline\hline  
Ion &  $\mathrm{\lambda_{rest}}$ & $\mathrm{\lambda_{obs}}$ & FWHM                         &   \multicolumn{1}{c}{Measured Flux}      \\ 
    &  ($\mathrm{\AA}$)         & ($\mathrm{\AA}$)         & ($\mathrm{km \cdot s^{-1}}$)  &   \multicolumn{1}{c}{($\mathrm{10^{-17} \cdot erg s^{-1}\cdot cm^{-2}}$)}      \\ 
\hline 
                                                                                                        
Ly$\alpha$      &      1215.7                      &      5010.1 $\pm$ 0.2     & 1848 $\pm$ 70 & 69     ,  4   \\
NV              &      1238.8, 1242.8 (1240.15)    &      5107.3 $\pm$ 2.3     & -             & 8.2    ,  1.1    \\
SiIV$^{a}$      &      1393.7                      &      5723.0 $\pm$ 4.1     & -             & -8.2   ,  1.0   \\
SiIV            &      1042.8                      &      5770.9 $\pm$ 4.7     & -             & 2.8    ,  1.6  \\
CIV             &      1548.2, 1550.8 (1549.06)    &      6381.3 $\pm$ 1.4     & 1373 $\pm$ 61 & 12.9   ,  1.2   \\
HeII            &      1640.4                      &      6747.3 $\pm$ 5.3     & -             & 4.9    ,  0.9    \\
OIII]           &      1660.8                      &      6840.4 $\pm$ 2.7     & -             & 6.0    ,  0.8  \\
CIII]           &      1906.8,1908.7               &      7843.6 $\pm$ 1.4     & -             & 11.3   ,  1.4   \\     
 
\end{tabular}
\begin{flushleft}
$^a$ Absorption 
\end{flushleft}
\end{table*}

\end{appendix}
\end{document}